\pdfoutput=1

\newif\ifdraft\draftfalse  
\newif\iffull\fullfalse   
\newif\ifbackref\backreffalse 

\makeatletter \@input{texdirectives.tex} \makeatother

\PassOptionsToPackage{names,dvipsnames}{xcolor} 
\documentclass[sigplan,10pt]{acmart}

\setcopyright{none}             

\makeatletter
\fancypagestyle{firstpagestyle}{%
  \fancyhf{}%
  \renewcommand{\headrulewidth}{\z@}%
  \renewcommand{\footrulewidth}{\z@}%
    \fancyhead[L]{}%
    \fancyhead[R]{}%
}
\fancypagestyle{standardpagestyle}{%
  \fancyhf{}%
  \renewcommand{\headrulewidth}{\z@}%
  \renewcommand{\footrulewidth}{\z@}%
    \fancyhead[LE]{\@headfootfont\thepage}%
    \fancyhead[RO]{\@headfootfont\thepage}%
    \fancyhead[RE]{\@headfootfont\@shortauthors}%
    \fancyhead[LO]{\@headfootfont\shorttitle}%
    \fancyfoot[RO,LE]{}%
}
\def\@mkbibcitation{}
\makeatother

\renewcommand\footnotetextcopyrightpermission[1]{}

\usepackage[justification=centering]{caption}
\usepackage{afterpage}
\usepackage{color}
\usepackage{amssymb,amsmath,amsfonts}
\usepackage{extarrows}
\usepackage{version}
\usepackage{xspace}
\usepackage{graphicx}
\usepackage{listings}
\usepackage{wrapfig}
\usepackage{ifpdf}
\usepackage{semantic}
\usepackage{mathpartir} 
\usepackage{natbib}
\setcitestyle{numbers}
\usepackage{amsthm}
\usepackage{stmaryrd}
\usepackage{subfigure}
\usepackage{multirow}
\usepackage{proof}
\definecolor{darkblue}{rgb}{0.0,0.0,0.3}

\usepackage{cleveref}
\usepackage{color}
\usepackage{listings}
\usepackage[colorinlistoftodos]{todonotes}
\usepackage{tikz}
\usetikzlibrary{positioning,shadows,arrows,calc,backgrounds,fit,shapes,shapes.multipart,decorations.pathreplacing,shapes.misc,patterns}
\usepackage{tikzscale}
\usepackage{xspace}

\usepackage{hyperref}
\hypersetup{breaklinks=true}

\newcommand*{\ETAL}{et al.\xspace}

\begin{document}

\title{\huge Robust Hyperproperty Preservation for Secure Compilation}

\subtitle{(Extended Abstract)\vspace{-0.5em}}

\author{\normalsize
       Deepak Garg$^1$ \quad
       C\u{a}t\u{a}lin Hri\c{t}cu$^2$ \quad
       Marco Patrignani$^3$ \quad
       Marco Stronati$^2$ \quad
       David Swasey$^1$\vspace{0.5em}}
\affiliation{$^1$MPI-SWS \qquad
             $^2$Inria Paris \qquad
             $^3$CISPA}

\makeatletter
\renewcommand{\@shortauthors}{
Deepak Garg, 
C\u{a}t\u{a}lin Hri\c{t}cu,
Marco Patrignani,
Marco Stronati, and
David Swasey}
\renewcommand{\shorttitle}{Robust Hyperproperty Preservation
  for Secure Compilation}
\makeatother

\begin{abstract}
We map the space of soundness criteria for secure compilation based on
the preservation of hyperproperties in arbitrary adversarial contexts,
which we call robust hyperproperty preservation. For this, we study
the preservation of several classes of hyperproperties and for each
class we propose an equivalent "property-free" characterization of
secure compilation that is generally better tailored for proofs.
Even the strongest of our soundness criteria, the
robust preservation of all hyperproperties, seems achievable for
simple transformations and provable using context back-translation
techniques previously developed for showing fully abstract compilation.
While proving the robust preservation of hyperproperties that are not
safety requires such powerful context back-translation techniques, for
preserving safety hyperproperties robustly, translating each finite
trace prefix back to a source context seems to suffice.

\end{abstract}

\maketitle

\newcommand{\cmp}[1]{#1\hspace{-0.35em}\downarrow}

\section*{Extended Abstract}
\label{sec:intro}

Secure compilation is an emerging field that puts together advances in
programming languages, verification, compilers, and security
enforcement mechanisms to devise secure compiler chains that eliminate
many of today's devastating low-level vulnerabilities.
One class of low-level vulnerabilities arises when code written in a
safe language is compiled and interacts with unsafe code written in a
lower-level language, e.g., when linking with libraries.
While currently all the guarantees of the source code are generally
lost in such cases, we would like to devise secure compilers that
protect some of the security guarantees established in the source
language even against adversarial low-level contexts.

What is a good soundness criterion for a compiler that attains this?
Fully abstract compilation~\cite{Abadi99} is a criterion that provides
one potential answer to this question: a fully abstract compiler preserves (and reflects)
the observational equivalence of partial source
programs.
In more detail, a compiler is fully abstract when any two partial
source programs that are observationally indistinguishable by all
compatible adversarial source contexts get compiled to two target-level
programs that are indistinguishable by all adversarial target-level contexts.
While fully abstract compilation has received significant attention in
the literature, the indistinguishability of partial programs in all
contexts is not the only security property one might be interested in
preserving.
In this work we set out to explore a much larger space of security
properties that can be preserved even against adversarial target-level
contexts.

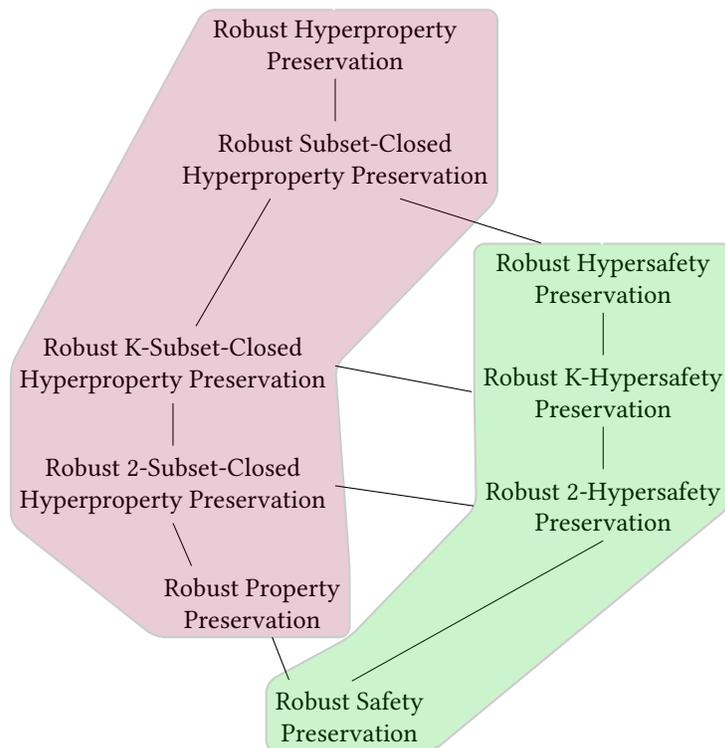
\begin{figure}[t!]
\vspace{1em}
\begin{tikzpicture}[shorten >=1pt,auto,node distance=6mm]
\tikzstyle{state} =[fill=white,minimum size=4pt]
\tikzstyle{field} =[fill=gray!5,draw=black!70, rectangle, minimum width={width("whiskersfieldww")+2pt}]]

\node[anchor=east, align=center] (h) { Robust Hyperproperty \\Preservation };

			\node[align=center, below = of h.south] (sc) { Robust Subset-Closed \\Hyperproperty Preservation };
				\node[align=center, below = of sc.south east, xshift = 4em] (hs) { Robust Hypersafety \\Preservation };
					\node[align=center, below = of hs.south] (khs) { Robust K-Hypersafety \\Preservation };
						\node[align=center, below = of khs.south] (2hs) { Robust 2-Hypersafety \\Preservation };


				\node[align=center, yshift = 1em] at ( sc.west |- khs) (ksc) { Robust K-Subset-Closed \\Hyperproperty Preservation };
					\node[align=center, below = of ksc.south] (2sc) { Robust 2-Subset-Closed \\Hyperproperty Preservation };
						\node[align=center, below = of 2sc.south, xshift = 3em] (p) { Robust Property \\Preservation };

							\node[align=center, below = of p.south east] (sp) { Robust Safety \\Preservation };
		
		\draw[-] (h.-90) to (sc.90);


		\draw[-] (sc.-150) to (ksc.60);
		\draw[-] (sc.-30) to (hs.150);

		\draw[-] (ksc.-90) to (2sc.90);
		
		\draw[-] (hs.-90) to (khs.90);

		\draw[-] (khs.-90) to (2hs.90);

		\draw[-] (ksc.0) to (khs.180);
		\draw[-] (2sc.0) to (2hs.180);

		\draw[-] (2sc.-90) to (p.150);

		\draw[-] (2hs.-90) to (sp.90);

		\draw[-] (p.-60) to (sp.150);

		\draw[rounded corners, thick, fill=green!80!black, opacity=0.2]
    	(hs.north) -| (khs.east) -- (2hs.east) -- (sp.south east) -| (sp.north west) -- (p.south east) -- (2hs.west) -- (khs.west) |- (hs.north);

    	\draw[rounded corners, thick, fill=purple!80!black, opacity=0.2]
    	(h.north) -| (sc.south east) -- (ksc.east) -- (p.north east) -- (p.south east) -- (p.south west) -- (2sc.south west) -- (ksc.west) -- (h.north west) -- (h.north);
\end{tikzpicture}
\caption{Different notions of robust hyperproperty preservation. Notions higher in the figure are stronger.}
\label{fig:notions}
\end{figure}

Specifically, we look at preserving classes of {\em
hyperproperties} despite adversarial contexts.
Hyperproperties~\cite{ClarksonS10} are a generalization of trace
properties that can express important security policies such as
noninterference.
While trace properties are formally expressed as sets of (potentially
infinite) traces, hyperproperties are sets of sets of traces.
Concretely, these traces are built over events such as inputs from and
outputs to the environment~\cite{Leroy09}.
We say that a complete program $P$ satisfies a hyperproperty $H$ when
the set of traces of $P$  is a member of $H$, or formally
$\{t ~|~ P \rightsquigarrow t\} \in H$, where $P \rightsquigarrow t$ indicates that program $P$ emits trace $t$.
We say that a partial program $P$ {\em robustly
  satisfies}~\cite{KupfermanV99} a hyperproperty $H$ when $P$
  linked with any (adversarial) context satisfies $H$.
Armed with this notion of robust satisfaction of hyperproperties, we
define secure compilation as preserving the robust satisfaction of a
class of hyperproperties $\mathcal{H}$, so if a partial source program
$P$ robustly satisfies a hyperproperty $H \in \mathcal{H}$ (wrt.\ all
source contexts) then its compilation $\cmp{P}$ must also robustly
satisfy $H$ (wrt.\ all target-level contexts).

We study the preservation of robust satisfaction for various classes
of hyperproperties, many of which are mentioned in
\Cref{fig:notions}, and which include all hyperproperties,
subset-closed hyperproperties, safety hyperproperties,
trace properties, and safety properties.
%
%
For each such class we propose an equivalent ``property-free''
characterization of secure compilation that is generally better
suited for proofs.
For instance, we prove that preserving all hyperproperties robustly
can be equivalently stated as the following criterion we call {\em
hyper-robust compilation} (where $C$ are contexts and $C[P]$ is the linking of a context $C$ with a program $P$):
\[
\forall P. \forall C_T. \exists C_S. \forall t. ~ C_T[\cmp{P}] \rightsquigarrow t \iff
                                       C_S[P] \rightsquigarrow t
\]
This requires that, given a program $P$, each target context $C_T$ can
be mapped to a source context $C_S$ in a way that perfectly preserves
the set of traces produced when linking with $P$ and $\cmp{P}$
respectively.
On the other hand, preserving all trace properties robustly is
equivalent to the following {\em robust compilation} criterion:
\[
\forall P. \forall C_T. \forall t. \exists C_S.~
  C_T[\cmp{P}] \rightsquigarrow t \Rightarrow  C_S[P] \rightsquigarrow t
\]
Compared to the previous definition, the $\exists C_S$ and $\forall t$
quantifiers in this definition are swapped and the implication is in
just one direction: Each (bad) trace in the target can be emulated
using a different source context $C_S$.
The intuition is that if the compiled program is able to produce a trace, that same trace must also be produceable in the source.
Swapping the quantifiers is crucial for transitioning from hyperproperties (sets of traces) to properties (traces).
The $\forall t. \exists C_S$ quantifiers of robust
compilation let us pick a different context $C_S$ for each trace $t$.
The reversed ordering $\exists C_S. \forall t$ of robust hyperproperty
preservation instead requires us pick the context $C_S$ before the
traces.
%
As a final example, preserving only safety properties robustly~\cite{GordonJ04} is equivalent to the following {\em robustly safe
  compilation} criterion:
\[
\forall P. \forall C_T. \forall t.~
  C_T[\cmp{P}] \rightsquigarrow t \Rightarrow
  \forall m {\leq} t.~ \exists C_S~t'.~  C_S[P] \rightsquigarrow t' \wedge m {\leq} t'
\]
Here only the (bad) finite prefixes $m$ of a potentially infinite
trace $t$ in the target need to be back-simulated in the source.
Safety properties are concerned with (bad) prefixes that must not
happen for the property to hold.
If a safety property holds in the source and some prefix broke this
property in the target, then the same prefix would also exist in the
source, contradicting the fact that the property holds in the source.

Even the strongest of our secure compilation criteria, hyper-robust
compilation, which is, as explained above, equivalent to the robust
preservation of all hyperproperties, seems achievable.
We plan to demonstrate this by adapting a recent fully
abstract translation of a simply typed $\lambda$-calculus into the
untyped $\lambda$-calculus~\cite{DevriesePP16}.
For this to be interesting, we first extend the two $\lambda$-calculi
with a notion of trace by adding inputs from and outputs to the
environment.
For achieving hyper-robust compilation we also extend the source
language with recursive types.
This allows us to encode the unitype of untyped $\lambda$-calculus
values using recursive, product, and sum types, allowing for a precise
back-translation of contexts.
We expect that the logical relation proof technique of Devriese
\ETAL~\cite{DevriesePP16} can be adapted to prove the hyper-robust
compilation of their translation.
Moreover, if we drop recursive types from the source we expect
to still be able to use the approximate back-translation of Devriese
\ETAL~\cite{DevriesePP16} to show robust hypersafety preservation, a
weaker security criterion.

While preserving hyperproperties that are not safety seems to require
powerful context back-translation techniques, for preserving safety
hyperproperties translating each finite trace prefix individually back
to a source context is also possible.
This could potentially be simpler as it can benefit from proof
techniques that are based on trace semantics~\cite{JeffreyR05},
which was also used in the context of full abstraction
proofs~\cite{PatrignaniC15, JuglaretHAEP16}.
In \Cref{fig:notions} we mark in green the secure compilation
criteria for which mapping finite trace prefixes is possible, and in
light purple the ones for which it is not.
%

Finally, the property-free characterizations of all classes of
hyperproperties from \Cref{fig:notions} have quantifier
alternation of the form $\forall P.\forall C_T \ldots \exists
C_S\ldots$, so a (constructive) proof that a compiler satisfies such a
characterization can            define $C_S$ as a function of the source program
$P$ and the target context $C_T$.
While the dependence of $C_S$ on $C_T$ is essential in most cases, the
dependence of $C_S$ on $P$ is necessary only when the target language
allows the context to make observations that the source does not allow.
For example, the target language may have reflection but the source
language may not have it.
However, in many cases, this kind of an abstraction mismatch does not
exist and, in fact, many existing proof techniques~\cite{NewBA16}, including the
aforementioned context back-translation techniques, construct $C_S$
only from $C_T$, independent of the source program $P$.
This begs the question of what kinds of properties are actually
preserved by a compiler that satisfies a \emph{stronger} criterion of
the form $\forall C_T. \exists C_S. \forall P \ldots$.
For example, what properties of source programs are preserved by a
compiler that satisfies the following stronger variant of
hyper-robust compilation?
\[
\forall C_T. \exists C_S. \forall P. \forall t. ~ C_T[\cmp{P}] \rightsquigarrow t \iff
                                       C_S[P] \rightsquigarrow t
\]
We conjecture that such strong soundness criteria correspond to the
robust preservation of \emph{relational} properties of programs.
In particular, the criterion listed above implies (the interesting
direction of) full abstraction, which
is the robust preservation of a specific relational property, namely,
observational equivalence.
%
%
Investigating which of the criteria of \Cref{fig:notions} imply or are
implied by fully abstract compilation is interesting future work.

\clearpage

\bibliographystyle{plainurl}
\bibliography{mp}

\end{document}